\documentclass[12pt]{article}
\begin{document}
\input{epsf}
\Large
\begin{center}
\bf{
The strange-sea quark spin distribution in the nucleon from 
inclusive and semi-inclusive deep-inelastic scattering
}
\end{center}

\normalsize
\begin{center}
      L.~Grigoryan
\end{center}

\begin{center}
      Yerevan Physics Institute, Br.Alikhanian 2, 375036 Yerevan, Armenia
\end{center}
\begin {abstract}
\hspace*{1em} 
We propose new method which allows determination of the strange-sea quark
spin distribution in the nucleon through measurement of various inclusive
and semi-inclusive polarized deep inelastic electron- or muon-proton reactions.
It is shown, that using combinations of inclusive data
and semi-inclusive data containing neutral pions in the final state, it is 
possible to extract the strange-sea quark spin distribution.
Similar result can be obtained for charged pions and some other hadrons also.
\end {abstract} 
\normalsize
\hspace*{1em} 
The standard parton model (see, for instance~\cite{feynman,close}) allows
to study spin structure of nucleons using both inclusive and semi-inclusive
polarized deep-inelastic scattering (DIS) of leptons on nucleons. 
According to this model,
an incoming electron or muon emits a spacelike virtual photon, which is
absorbed by a quark in the nucleon. The nucleon is broken up, and the struck
quark and the target remnant fragment into hadrons in the final state.
The pioneering experiments to study the spin structure of the nucleon
performed at SLAC~\cite{alguard,baum} were measurements of inclusive spin
asymmetries, in which, in final state, only the scattered lepton is observed.
More precise later experiments (see, for instance~\cite{ashman}) showed that
only a fraction of the nucleon spin can be attributed to the quark spins and
that the strange quark sea seems to be negatively polarized~\cite{ellis}.
The polarized semi-inclusive DIS with detection of one hadron in the final 
state adds important information on the scattering process.
At present there are
many theoretical and experimental efforts devoted the study of the valence and
sea quarks spin distributions in the nucleon. Here we are interested in
the study of the strange-sea quark spin distribution. 
This point was already investigated earlier. Two different methods 
for the extraction of strange-sea quark spin distribution
were proposed~\cite{frankfurt,hermes1}.
One of them was applied to the experimental data~\cite{hermes1,hermes2,hermes3}.\\
\hspace*{1em}
The objective of this work is to propose new method for the extraction,
which, in our opinion, is simpler than preceding methods and, 
what is more important, allows to extract the same information using other
set of experimental semi-inclusive DIS data. Together with the other available 
methods, it can be used by experimentalists for more detailed investigation of 
the subject.\\
\hspace*{1em} 
Let us consider at first the inclusive DIS 
of lepton on proton $l_{i} + p \to l_{f} + X$, where $l_{i}(l_{f})$ are initial
(final) leptons, $p$ is a proton and $X$ is undetected state. Reaction can be
wrote directly for intermediate virtual photon, as $\gamma^{*} + p \to X$.
We are interested the number of inclusive DIS events
$N^{inc}$ produced in a given bin of Bjorken variable $x = Q^{2}/2M_{p}\nu$, 
where $q^{2} = -Q^{2}$ and $\nu$ are the square of the four-momentum and energy 
of the virtual photon, $M_{p}$ is the proton mass, in configurations where the
virtual photon and target proton helicities are antiparallel
$(\uparrow \downarrow)$ or parallel $(\uparrow \uparrow)$. The standard parton 
model consideration~\cite{feynman,close} gives, up to constant factor:
\begin{eqnarray}
N^{inc}_{\uparrow \downarrow} \sim \frac{4}{9}u_{+}(x) +
\frac{4}{9}\bar{u}_{+}(x) +
\frac{1}{9}d_{+}(x) +
\frac{1}{9}\bar{d}_{+}(x)
+ \frac{1}{9}s_{+}(x) +
\frac{1}{9}\bar{s}_{+}(x) ,
\end{eqnarray}
\begin{eqnarray}
N^{inc}_{\uparrow \uparrow} \sim \frac{4}{9}u_{-}(x) +
\frac{4}{9}\bar{u}_{-}(x) +
\frac{1}{9}d_{-}(x) +
\frac{1}{9}\bar{d}_{-}(x)
+ \frac{1}{9}s_{-}(x) +
\frac{1}{9}\bar{s}_{-}(x) ,
\end{eqnarray}
where $q_{+} (q_{-})$ denotes distribution function of a quark $q$ with helicity
parallel (antiparallel) to the proton helicity.\\
\hspace*{1em}
Now let us turn to the semi-inclusive DIS reaction with detection in final state
$\pi^0$ meson in coincidence with the scattered lepton 
$l_{i} + p \to l_{f} + \pi^0 + X$, 
or for intermediate virtual photon, as $\gamma^{*} + p \to \pi^0 + X$.
We will consider the number of $\pi^{0}$ mesons $N^{\pi^{0}}$ produced in 
a given bin of variable $x$ and $z$ ($z = E_{h}/\nu$, where $E_{h}$ is the final
pion energy in lab. system), which is the fraction of the virtual photon
energy taken by the pion, in configuration where the virtual photon and target 
proton helicities are antiparallel $(\uparrow \downarrow)$ or parallel 
$(\uparrow \uparrow)$. The standard parton model 
consideration~\cite{feynman,close} gives, up to constant factor:
\begin{eqnarray}
N^{\pi^{0}}_{\uparrow \downarrow} \sim \frac{4}{9}u_{+}(x)D^{\pi^{0}}_{u}(z) +
\frac{4}{9}\bar{u}_{+}(x)D^{\pi^{0}}_{\bar{u}}(z) +
\frac{1}{9}d_{+}(x)D^{\pi^{0}}_{d}(z) +
\frac{1}{9}\bar{d}_{+}(x)D^{\pi^{0}}_{\bar{d}}(z) 
\nonumber
\end{eqnarray}
\begin{eqnarray}
+ \frac{1}{9}s_{+}(x)D^{\pi^{0}}_{s}(z) +
\frac{1}{9}\bar{s}_{+}(x)D^{\pi^{0}}_{\bar{s}}(z) ,
\end{eqnarray}
\begin{eqnarray}
N^{\pi^{0}}_{\uparrow \uparrow} \sim \frac{4}{9}u_{-}(x)D^{\pi^{0}}_{u}(z) +
\frac{4}{9}\bar{u}_{-}(x)D^{\pi^{0}}_{\bar{u}}(z) +
\frac{1}{9}d_{-}(x)D^{\pi^{0}}_{d}(z) +
\frac{1}{9}\bar{d}_{-}(x)D^{\pi^{0}}_{\bar{d}}(z)
\nonumber
\end{eqnarray}
\begin{eqnarray}
+ \frac{1}{9}s_{-}(x)D^{\pi^{0}}_{s}(z) +
\frac{1}{9}\bar{s}_{-}(x)D^{\pi^{0}}_{\bar{s}}(z) ,
\end{eqnarray}
where $D^{h}_{q}(z)$ is the fragmentation
function of a quark $q$ into the hadron $h$ with energy $E_{h} = z \cdot \nu$. 
We will follow the notations and approximations for fragmentation functions,
which were used in~\cite{frankfurt}. 
The above relations have been obtained with the standard assumption that the 
fragmentation function $D^{h}_{q}(z)$ does not depend on the quark helicity i.e., 
that $D^{h}_{q_{+}}(z) = D^{h}_{q_{-}}(z)$. Now we may use the isospin and charge 
conjugation symmetry to relate various fragmentation functions, so finally we are 
left, for case of charged pions, with three independent fragmentation functions 
$D_{1}(z)$, $D_{2}(z)$ and $D_{3}(z)$, called favoured, unfavoured and strange 
quark fragmentation functions, correspondingly. 
\begin{eqnarray}
D_{1}(z) \equiv
D^{\pi^{+}}_{u}(z) = D^{\pi^{-}}_{\bar{u}}(z) =D^{\pi^{-}}_{d}(z) =
D^{\pi^{+}}_{\bar{d}}(z) ,
\end{eqnarray}  
\begin{eqnarray}
D_{2}(z) \equiv
D^{\pi^{-}}_{u}(z) = D^{\pi^{+}}_{\bar{u}}(z) =D^{\pi^{+}}_{d}(z) =
D^{\pi^{-}}_{\bar{d}}(z) ,
\end{eqnarray}
\begin{eqnarray}
D_{3}(z) \equiv D^{\pi^{+}}_{s}(z) = D^{\pi^{+}}_{\bar{s}}(z) =
D^{\pi^{-}}_{s}(z) = D^{\pi^{-}}_{\bar{s}}(z).
\end{eqnarray}
In case of neutral pion situation is simpler, because the wave function of the 
$\pi^{0}$ meson is:
\begin{eqnarray}
\pi^{0} = \frac{u\bar{u} - d\bar{d}}{\sqrt{2}} ,
\end{eqnarray}
and we are left with two fragmentation functions 
$D_{0}(z) = D_{1}(z) + D_{2}(z)$ and $D_{3}(z)$.
\begin{eqnarray}
\frac{1}{2}D_{0}(z) \equiv
D^{\pi^{0}}_{u}(z) = D^{\pi^{0}}_{\bar{u}}(z) =D^{\pi^{0}}_{d}(z) =
D^{\pi^{0}}_{\bar{d}}(z) ,
\end{eqnarray}
\begin{eqnarray}
D_{3}(z) \equiv D^{\pi^{0}}_{s}(z) = D^{\pi^{0}}_{\bar{s}}(z)
\end{eqnarray}
In this notations, eqs. (3)-(4) take the form
\begin{eqnarray}
N^{\pi^{0}}_{\uparrow \downarrow} \sim [\frac{4}{9}u_{+}(x) +
\frac{4}{9}\bar{u}_{+}(x) +
\frac{1}{9}d_{+}(x) +
\frac{1}{9}\bar{d}_{+}(x)]\frac{1}{2}D_{0}(z)
+ [\frac{1}{9}s_{+}(x) +
\frac{1}{9}\bar{s}_{+}(x)]D_{3}(z) ,
\end{eqnarray}
\begin{eqnarray}
N^{\pi^{0}}_{\uparrow \uparrow} \sim [\frac{4}{9}u_{-}(x) +
\frac{4}{9}\bar{u}_{-}(x) +
\frac{1}{9}d_{-}(x) +
\frac{1}{9}\bar{d}_{-}(x)]\frac{1}{2}D_{0}(z)
+ [\frac{1}{9}s_{-}(x) +
\frac{1}{9}\bar{s}_{-}(x)]D_{3}(z) .
\end{eqnarray}
It is convenient instead of $q_{+}(x)$ and $q_{-}(x)$ introduce the standard
quark distribution functions $q(x)$ and $\Delta q(x)$ :
\begin{eqnarray}
q(x) \equiv q_{+}(x) + q_{-}(x), \hspace{1cm} \Delta q(x) \equiv q_{+}(x) - 
q_{-}(x) .
\end{eqnarray}
Then from eqs. (1)-(2) and (11)-(12) follows:
\begin{eqnarray}
N^{inc}_{\uparrow \downarrow} - N^{inc}_{\uparrow \uparrow}
\sim \frac{4}{9}\Delta u(x) + \frac{4}{9}\Delta \bar{u}(x) +
\frac{1}{9}\Delta d(x) + \frac{1}{9}\Delta \bar{d}(x)
+ \frac{1}{9}\Delta s(x) + \frac{1}{9}\Delta \bar{s}(x) ,
\end{eqnarray}
\begin{eqnarray}
N^{inc}_{\uparrow \downarrow} + N^{inc}_{\uparrow \uparrow}
\sim \frac{4}{9}u(x) + \frac{4}{9}\bar{u}(x) + \frac{1}{9}d(x) +
\frac{1}{9}\bar{d}(x) + \frac{1}{9}s(x) + \frac{1}{9}\bar{s}(x) ,
\end{eqnarray}
\begin{eqnarray}
N^{\pi^{0}}_{\uparrow \downarrow} - N^{\pi^{0}}_{\uparrow \uparrow}
\sim [\frac{4}{9}\Delta u(x) +
\frac{4}{9}\Delta \bar{u}(x) +
\frac{1}{9}\Delta d(x) +
\frac{1}{9}\Delta \bar{d}(x)]\frac{1}{2}D_{0}(z)
\nonumber
\end{eqnarray}
\begin{eqnarray}
+ [\frac{1}{9}\Delta s(x) +
\frac{1}{9}\Delta \bar{s}(x)]D_{3}(z) ,
\end{eqnarray}  
\begin{eqnarray}
N^{\pi^{0}}_{\uparrow \downarrow} + N^{\pi^{0}}_{\uparrow \uparrow}
\sim [\frac{4}{9}u(x) +
\frac{4}{9}\bar{u}(x) +
\frac{1}{9}d(x) +
\frac{1}{9}\bar{d}(x)]\frac{1}{2}D_{0}(z)
+ [\frac{1}{9}s(x) +
\frac{1}{9}\bar{s}(x)]D_{3}(z) .
\end{eqnarray}
Now we can obtain expressions for strange sea polarization, using two ways.\\
\hspace*{1em}
First way includes combinations of inclusive and semi-inclusive data. 
From eqs. (14)-(17) we obtain for strange sea polarization:
\begin{eqnarray}
\frac{\Delta s(x) + \Delta \bar{s}(x)}{s(x) + \bar{s}(x)} =
\frac{N^{\pi^{0}}_{\uparrow \downarrow} - N^{\pi^{0}}_{\uparrow \uparrow}
 - \frac{1}{2}D_{0}(z)(N^{inc}_{\uparrow \downarrow} - N^{inc}_{\uparrow 
\uparrow})}
{N^{\pi^{0}}_{\uparrow \downarrow} + N^{\pi^{0}}_{\uparrow \uparrow}
 - \frac{1}{2}D_{0}(z)(N^{inc}_{\uparrow \downarrow} + N^{inc}_{\uparrow 
\uparrow})} .
\end{eqnarray}
For increasing of statistics it is desirable take semi-inclusive data in 
maximal wide region of $z$, $z_{min} < z < z_{max}$, then instead of 
fragmentation function $D_{0}(z)$ one need to take integral from it over $z$.
Eq.(18) can be written in the simple form $(1 - r)/(1 + r)$,
where $r = (N^{\pi^{0}}_{\uparrow \uparrow} - \frac{1}{2}D_{0}(z)N^{inc}_{\uparrow 
\uparrow})/
(N^{\pi^{0}}_{\uparrow \downarrow} - \frac{1}{2}D_{0}(z)N^{inc}_{\uparrow  
\downarrow})$.\\
\hspace*{1em}
Second way includes semi-inclusive $\pi^{0}$ meson production data only and can
be expressed in form:
\begin{eqnarray}
\frac{\Delta s(x) + \Delta \bar{s}(x)}{s(x) + \bar{s}(x)} =
\frac{(N^{\pi^{0}}_{\uparrow \downarrow} - N^{\pi^{0}}_{\uparrow \uparrow})_{1} -
R_{12}(N^{\pi^{0}}_{\uparrow \downarrow} - N^{\pi^{0}}_{\uparrow \uparrow})_{2}}
{(N^{\pi^{0}}_{\uparrow \downarrow} + N^{\pi^{0}}_{\uparrow \uparrow})_{1} - 
R_{12}(N^{\pi^{0}}_{\uparrow \downarrow} + N^{\pi^{0}}_{\uparrow \uparrow})_{2}} ,
\end{eqnarray}
where $(N^{\pi^{0}}_{\uparrow \downarrow} \pm N^{\pi^{0}}_{\uparrow \uparrow})_{1}$
$((N^{\pi^{0}}_{\uparrow \downarrow} \pm N^{\pi^{0}}_{\uparrow \uparrow})_{2})$
means, that corresponding data taken in region $z_{1}$ ($z_{2}$).
Here $z_{1}$ and $z_{2}$ are two different bins on $z$. 
It is desirable to take them maximal large as was discussed above. Factor $R_{12}$ 
is a ratio of fragmentation functions
in different regions of $z$ $R_{12} = D_{0}(z_{1})/D_{0}(z_{2})$.
Eq.(19) also can be written in the form $(1 - r)/(1 + r)$,
with $r = (N^{\pi^{0}}_{\uparrow \uparrow 1} - N^{\pi^{0}}_{\uparrow \uparrow 2} R_{12}
)/
(N^{\pi^{0}}_{\uparrow \downarrow 1} - N^{\pi^{0}}_{\uparrow \downarrow 2} R_{12}
)$.\\
The above consideration can be repeated for combination of charged pions, 
$( \pi^{\pm} \equiv \pi^{+} + \pi^{-})$,
which finally gives instead of eqs. (18) and (19) similar equations for charged 
pions:
\begin{eqnarray}
\frac{\Delta s(x) + \Delta \bar{s}(x)}{s(x) + \bar{s}(x)} =
\frac{N^{\pi^{\pm}}_{\uparrow \downarrow} - N^{\pi^{\pm}}_{\uparrow \uparrow} -
D_{0}(z)(N^{inc}_{\uparrow \downarrow} - N^{inc}_{\uparrow 
\uparrow})}
{N^{\pi^{\pm}}_{\uparrow \downarrow} + N^{\pi^{\pm}}_{\uparrow \uparrow} -
D_{0}(z)(N^{inc}_{\uparrow \downarrow} + N^{inc}_{\uparrow 
\uparrow})}  ,
\end{eqnarray}
\begin{eqnarray}
\frac{\Delta s(x) + \Delta \bar{s}(x)}{s(x) + \bar{s}(x)} =
\frac{(N^{\pi^{\pm}}_{\uparrow \downarrow} - N^{\pi^{\pm}}_{\uparrow \uparrow})_{1} 
- R_{12}
(N^{\pi^{\pm}}_{\uparrow \downarrow} - N^{\pi^{\pm}}_{\uparrow \uparrow})_{2}}
{(N^{\pi^{\pm}}_{\uparrow \downarrow} + N^{\pi^{\pm}}_{\uparrow \uparrow})_{1}
- R_{12}
(N^{\pi^{\pm}}_{\uparrow \downarrow} + N^{\pi^{\pm}}_{\uparrow \uparrow})_{2}} .
\end{eqnarray}
Let us briefly discuss eqs.(18)-(21), which are main results of this study.
Eq.(18) contains two combinations of inclusive as well as semi-inclusive
DIS polarized events and the combination of the so called favoured and
unfavoured fragmentation functions. The events
can be obtained from the same measurement. Fragmentation functions can
be obtained from DIS on nucleon or from $e^{+}e^{-}$ annihilation. Equation (19)
contains only two combinations of semi-inclusive DIS polarized events and the ratio 
of fragmentation functions measured in different bins of $z$. In our
opinion, this structure leads to the reduction of systematic errors in eq.(19) 
in comparison with eq.(18) and consequently
gives it some advantage. Equations (20)-(21) can be used
instead of eqs.(18)-(19) if arise problems with statistics.\\
This scheme without any change can be applied for the consideration of
$\rho^{0}$ mesons in final state also. 
Diffractive $\rho^{0}$ mesons can be rejected if to exclude from consideration
values of variable $z$ close to the unity.
If to suppose that $\phi$ meson is pure $s \bar{s}$ state, then expressions of type
eqs. (18) and (19) can be obtained for $\phi$ meson by changing:
\begin{eqnarray}
\pi^{0} \rightarrow \phi ; \hspace{1cm} \frac{1}{2}D_{0}(z) \rightarrow 
D^{\phi}_{u}(z) ,
\end{eqnarray}
where $D^{\phi}_{u}(z)$ is the fragmentation function for the $u$-quark to turn
into $\phi$ meson, and it is supposed also that $D^{\phi}_{u}(z) = D^{\phi}_{d}(z) =
D^{\phi}_{\bar{u}}(z) = D^{\phi}_{\bar{d}}(z)$.
Our final results (eqs. (18)-(21))
do not depend from the type of target and can be applied in the
same form for neutron and deuteron targets also.
As a last remark, one should note that the experimental 
realization of this scheme can be performed by HERMES experiment at DESY,
the COMPASS collaboration, EMC's successor at CERN and at JLAB after 
beam-energy upgrade.

\end{document}